\batchmode
\makeatletter
\def\input@path{{/Users/ananyabalakrishna/Documents/Disemmination/Writing/Arxived/4_Energy_harvester/Lyx/}}
\makeatother
\documentclass[10pt,twocolumn]{article}
\usepackage[latin9]{inputenc}
\usepackage[letterpaper]{geometry}
\usepackage{color}
\usepackage{array}
\usepackage{verbatim}
\usepackage{booktabs}
\usepackage{amsmath}
\usepackage{amssymb}
\usepackage{graphicx}
\usepackage{setspace}
\usepackage[unicode=true,pdfusetitle,
 bookmarks=true,bookmarksnumbered=false,bookmarksopen=false,
 breaklinks=false,pdfborder={0 0 1},backref=false,colorlinks=false]
 {hyperref}

\makeatletter

\providecommand{\tabularnewline}{\\}

\usepackage{ragged2e}
\usepackage{booktabs}
\usepackage{authblk}
\usepackage{setspace}
\usepackage{titlesec}
\titlespacing*{\section}{0pt}{1.1\baselineskip}{\baselineskip}

\usepackage[parfill]{parskip}
\usepackage{graphicx}
\usepackage{cite}
\usepackage{notoccite}

\title{Nanoscale domain patterns and a concept for an energy harvester}
\author{Ananya Renuka Balakrishna, John E. Huber}
\affil{Department of Engineering Science, University of Oxford, Oxford, OX1 3PJ, United Kingdom}

\date{}                    

\makeatother

\begin{document}
\twocolumn[   
\begin{@twocolumnfalse}  
\begin{center}
\textbf{\LARGE{}Nanoscale domain patterns and a concept for an energy
harvester}
\par\end{center}{\LARGE \par}

\bigskip{}

\begin{center}
Ananya Renuka Balakrishna, John E. Huber
\par\end{center}

\bigskip{}

\begin{center}
{\footnotesize{}Department of Engineering Science, University of Oxford,
Parks Road, Oxford OX1 3PJ, England, United Kingdom}
\par\end{center}{\footnotesize \par}

\smallskip{}

\begin{center}
{\scriptsize{}$^{*}$Email: john.huber@eng.ox.ac.uk}
\par\end{center}{\scriptsize \par}

\bigskip{}

\bigskip{}

\begin{abstract}
\textcolor{black}{The current work employs a phase-field model to
test the stability of nanoscale periodic domain patterns, and to explore
the application of one pattern in an energy harvester device. At first,
the stability of several periodic domain patterns with in-plane polarizations
is tested under stress-free and electric field-free conditions. It
is found that simple domain patterns with stripe-like features are
stable, while patterns with more complex domain configurations are
typically unstable at the nanoscale. Upon identifying a stable domain
pattern with suitable properties, a conceptual design of a thin film
energy harvester device is explored. The harvester is modelled as
a thin ferroelectric film bound to a substrate. In the initial state
a periodic stripe domain pattern with zero net charge on the top electrode
is modelled. On bending the substrate, a mechanical strain is induced
in the film, causing polarized domains to undergo ferroelectric switching
and thus generate electrical energy. The results demonstrate the working
cycle of a conceptual energy harvester which, on operating at kHz
frequencies, such as from vibrations in the environment, could produce
an area power density of about 40W/m$^{2}$. }

\medskip{}

\end{abstract}
\bigskip{}

\bigskip{}

\end{@twocolumnfalse} ]
\begin{singlespace}

\section*{{\small{}Introduction}}
\end{singlespace}

\begin{singlespace}
\noindent \textcolor{black}{\small{}Ferroelectric materials possess
spontaneous polarization in unit cells that can be reoriented by the
application of an electric field or mechanical stresses \cite{key-1}.
This reorientation of polarization is typically accompanied by a change
in surface charge distribution that finds applications in sensors
\cite{key-2,key-3}, energy harvesters \cite{key-4,key-5,key-6} and
memory elements \cite{key-7,key-8}. A typical ferroelectric contains
nanoscale regions of uniform polarization referred to as domains,
which fit compatibly together to form polarization patterns \cite{key-9,key-10,key-11}.
These polarization patterns are important to the working of nanoscale
or thin film based devices where the length scale is comparable to
the domain size \cite{key-3,key-12,key-13,key-14}, and the domain
configurations can be engineered to improve ferroelectric properties
\cite{key-15,key-16}. }{\small \par}
\end{singlespace}

\textcolor{black}{\small{}Several designs of nanoscale ferroelectric
energy harvesters are proposed that generate sizeable amount of power
\cite{key-5,key-17,key-18}. However, these devices typically contain
nanoscale features which pose fabrication difficulties \cite{key-5,key-17}
or require a complex configuration of domains in the initial state
\cite{key-5}. Here, we model an energy harvester comprising a monocrystalline
ferroelectric thin film bound to conductive substrate, a relatively
simple configuration to manufacture. With technological advances in
thin film fabrication \cite{key-19,key-20,key-21}, film thickness
of the order of $10\mathrm{nm}$ nanometres are achievable \cite{key-19,key-22,key-23}.
In the present work we consider a 40nm film with a simple alternating
configuration of domains that is commonly found in thin films \cite{key-24,key-25,key-26}. }{\small \par}

\textcolor{black}{\small{}Phase-field models have been employed to
simulate nanoscale properties of ferroelectrics \cite{key-27,key-28},
and ferromagnetics \cite{key-29} as well as to explore their application
in nanoscale devices \cite{key-27,key-30,key-31}. They offer great
advantages in tracking domain walls \cite{key-28}. Baek }\textit{\textcolor{black}{\small{}et
al.}}\textcolor{black}{\small{} \cite{key-29} used phase-field simulations
to demonstrate the stabilisation of ferroelastic switching in BiFeO$_{3}$,
which has applications in non-volatile magnetoelectric devices. Similarly
Balke }\textit{\textcolor{black}{\small{}et al.}}\textcolor{black}{\small{}
\cite{key-27}, employed phase-field simulations to illustrate enhanced
electric conductivity at BiFeO$_{3}$ vortex cores, thereby indicating
their potential application in integrated oxide electronic devices.
Here, we employ a phase-field model \cite{key-32,key-33} that has
been established for studying the behaviour of barium titanate. }{\small \par}

\textcolor{black}{\small{}In this paper, we first use a phase-field
model to assess several nanoscale periodic polarization patterns and
then explore a conceptual design of an energy harvester. Tsou }\textit{\textcolor{black}{\small{}et
al.}}\textcolor{black}{\small{} \cite{key-34} identified several
periodic polarization patterns that can form in tetragonal ferroelectrics
and possess useful functional properties. However, on testing the
stability of these polarization patterns it is found that most become
unstable when the domains are a few nanometres in size. By contrast,
a simpler, layered pattern is found to be stable, and allows the domain
walls to sweep back and forth under external loads. Therefore, we
explore the potential application of this pattern in an energy harvester
device. Here, a thin film of BaTiO$_{3}$ with stripe domain pattern
is modelled with plane strain and plane electric field conditions
on a conductive substrate. Mechanical loads are applied to the film
by substrate bending that causes polarization switching in the stripe
domain pattern. Using phase-field simulations we demonstrate the working
cycle of the energy harvester device and find a potential area power
density of up to $40\mathrm{W/m^{2}}$, at an operating frequency
of 1kHz. This is comparable with the power density of currently available
photovoltaic cells, and greater than typical piezoelectric vibration
harvesters, making the concept attractive for further development. }{\small \par}
\begin{singlespace}

\section*{{\small{}Phase-field model}}
\end{singlespace}

\noindent \textcolor{black}{\small{}We give only the briefest review
of the phase-field model used for the current study, as its details
have been reported elsewhere, by Landis and co-workers \cite{key-32,key-33}.
In this model, the volumetric Helmholtz free energy, $\psi$ is described
by:}{\small \par}

\noindent \textcolor{black}{\footnotesize{}
\begin{align}
\psi & =\frac{1}{2}a_{ijkl}P_{i,j}P_{k,l}+\frac{1}{2}\overline{a}_{ij}P_{i}P_{j}+\frac{1}{4}\overline{\overline{a}}_{ijkl}P_{i}P_{j}P_{k}P_{l}\nonumber \\
 & ~+\frac{1}{6}\overline{\overline{\overline{a}}}_{ijklmn}P_{i}P_{j}P_{k}P_{l}P_{m}P_{n}+\frac{1}{8}\overline{\overline{\overline{\overline{a}}}}_{ijklmnrs}P_{i}P_{j}P_{k}P_{l}P_{m}P_{n}P_{r}P_{s}\nonumber \\
 & ~+b_{ijkl}\epsilon_{ij}P_{k}P_{l}+\frac{1}{2}c_{ijkl}\epsilon_{ij}\epsilon_{kl}+\frac{1}{2}f_{ijklmn}\epsilon_{ij}\epsilon_{kl}P_{m}P_{n}\nonumber \\
 & \ +\frac{1}{2}g_{ijklmn}\epsilon_{ij}P_{k}P_{l}P_{m}P_{n}+\frac{1}{2\kappa_{0}}(D_{i}-P_{i})(D_{i}-P_{i})\label{eq:1}
\end{align}
}{\footnotesize \par}

\textcolor{black}{\small{}Here $\psi$ is a function of polarization
$P_{i}$, polarization gradient $P_{i,j},$ strain $\epsilon_{ij}$
and electric displacement $D_{i}$. In equation \ref{eq:1} the first
term accounts for the domain wall energym while the remaining terms
contribute to the bulk energy of the ferroelectric system. Further
details on the tensorial coefficients $\mathbf{a},\mathbf{\bar{a}\ldots}\bar{\mathbf{\bar{\bar{\bar{a}}},}\mathbf{b,c,f}}$and
$\mathbf{g}$, and the material properties can be found in refs. \cite{key-32,key-33}.
The model is calibrated for barium titanate and is solved for equilibrium
using the time dependent Ginzburg-Landau equation \cite{key-32}: }{\small \par}

\noindent \textcolor{black}{\small{}
\begin{equation}
\left(\frac{\partial\psi}{\partial P_{i,j}}\right)_{,j}-\frac{\partial\psi}{\partial P_{i}}=\beta\dot{P_{i}},\label{eq:2}
\end{equation}
}{\small \par}

\textcolor{black}{\small{}The relaxation parameter $\beta$ is used
to find equilibrium states, which are then confirmed by solving the
system equations with $\beta=0$. This phase-field model is first
used to test the stability of several periodic domain patterns with
a view to identifying stable domain patterns with useful energy harvesting
properties. }{\small \par}
\begin{singlespace}

\section*{{\small{}Periodic domain patterns and their stability}}
\end{singlespace}

\textcolor{black}{\small{}The objective is to identify a periodic
pattern of domains that has useful properties for energy harvesting.
This is achieved by first testing the stability of various known periodic
domain patterns under conditions of zero average stress and electric
field to see which, if any, are likely to be usable at the nanoscale.
The candidate patterns are next tested under cyclic straining to see
which remain stable and can provide a mechanically induced cyclic
switching effect. }{\small \par}

\textcolor{black}{\small{}Planar domain patterns are modelled in 2-dimensions
on a periodic square region of side $L~=~40$nm, with plane strain
and plane electric field conditions. The imposed boundary conditions
allow the model to adopt states that are free of average stress or
electric field. This is achieved by relating the displacement, electric
potential and polarization values on typical boundary nodes indicated
in figure \ref{fig:1}. }{\small \par}
\begin{center}
{\footnotesize{}}
\begin{figure}
\begin{centering}
\includegraphics[width=0.4\columnwidth]{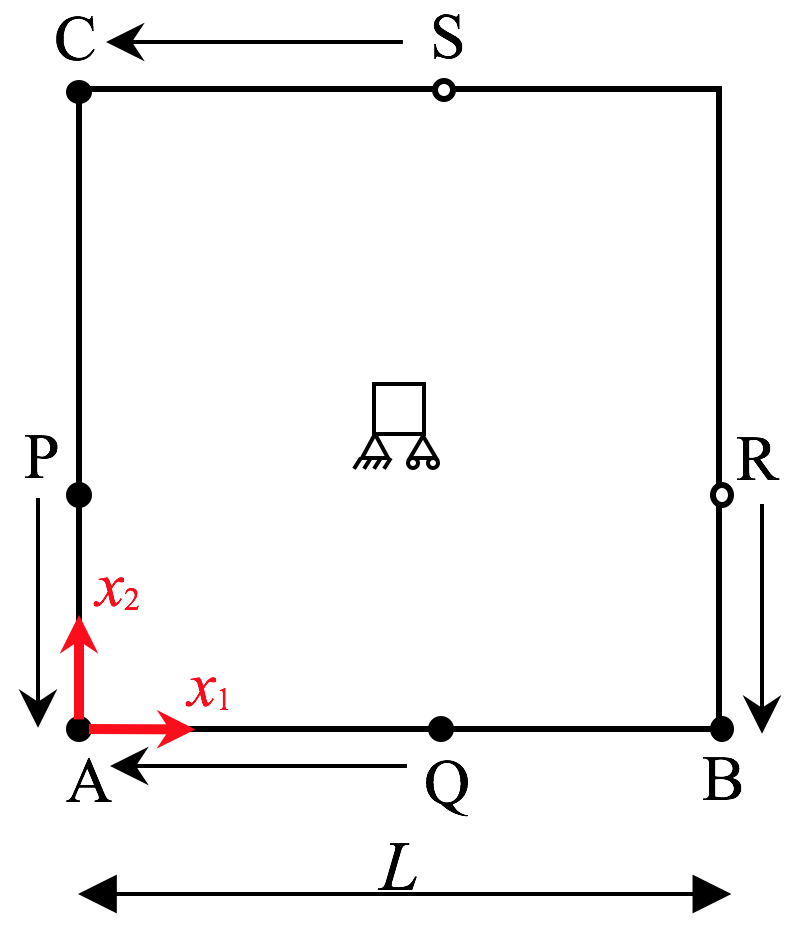}
\par\end{centering}
{\footnotesize{}\caption{{\small{}\label{fig:1}Schematic representation of boundary conditions
on square region barium titanate of side, $L$. Arrows indicate nodal
connectivity in periodic boundary conditions in equation \ref{eq:3}. }}
}{\footnotesize \par}
\end{figure}
\par\end{center}{\footnotesize \par}

\textcolor{black}{\small{}Referring to figure \ref{fig:1}, a mid-element
of the periodic square is simply supported and the boundary nodes
are modelled with periodic conditions. The boundary nodes are of two
types, namely master nodes (such as P on edge AC or Q on edge AB)
and dependent nodes (represented by R and S). }{\small \par}

\textcolor{black}{\small{}A representative example of the periodic
boundary conditions on the dependent nodes is:}{\small \par}

\noindent \textcolor{black}{\footnotesize{}
\begin{align}
u_{i}(L,x_{2})\mid_{R}-u_{i}(L,0)\mid_{\mathrm{B}} & =u_{i}(0,x_{2})\mid_{P}-u_{i}(0,0)\mid_{\mathrm{A}}\nonumber \\
\phi(L,x_{2})\mid_{R} & =\phi(0,x_{2})\mid_{P}\nonumber \\
P_{i}(L,x_{2})\mid_{R} & =P_{i}(0,x_{2})\mid_{P}\label{eq:3}
\end{align}
}{\footnotesize \par}

\noindent \textcolor{black}{\small{}where $u_{i}$ is the displacement,
$\phi$ is the electric potential and $P_{i}$ is the polarization
at each node. A similar set of periodic boundary condition is applied
on the representative node \textquoteleft S\textquoteright{} \cite{key-35}.
These constraints enforce periodicity of the strain fields with zero
average stress, and periodicity of the electric potential values with
zero average electric field. Periodicity of polarization is also imposed
at the boundary nodes. The boundary conditions given by equation }{\small{}\ref{eq:3}}\textcolor{black}{\small{}
are referred in this paper as the zero external load conditions. }{\small \par}

\textcolor{black}{\small{}In the initial state of each simulation,
polarization values corresponding to a known periodic pattern of domains
are imposed on each node of the periodic square with sharp domain
boundaries, see figure \ref{fig:2}(a-e). This initial state is next
allowed to relax towards equilibrium with slow stepping down of $\beta$
values to reach equilibrium at $\beta=0$. If the pattern persists
at equilibrium then it is stable, suggesting that the pattern could
form with domain wall spacings of order a few nm. }{\small \par}

\textcolor{black}{\small{}The stripe domain pattern, figure \ref{fig:2}(a)
and the herringbone domain pattern, figure \ref{fig:2}(b) are found
to be stable at equilibrium, while all the other patterns shown dissolved
into either single domain states or a simple stripe pattern. The energy
of each polarization pattern, and the associated strain and stress
fields have been discussed in previous work \cite{key-35} where it
was found that the stripe and herringbone pattern are stable due to
the absence of stressed domains. Each of the other patterns are stressed
due to incompatible domain junctions that increase their elastic energy. }{\small \par}
\begin{center}
{\footnotesize{}}
\begin{figure}
\begin{centering}
\includegraphics[width=1\columnwidth]{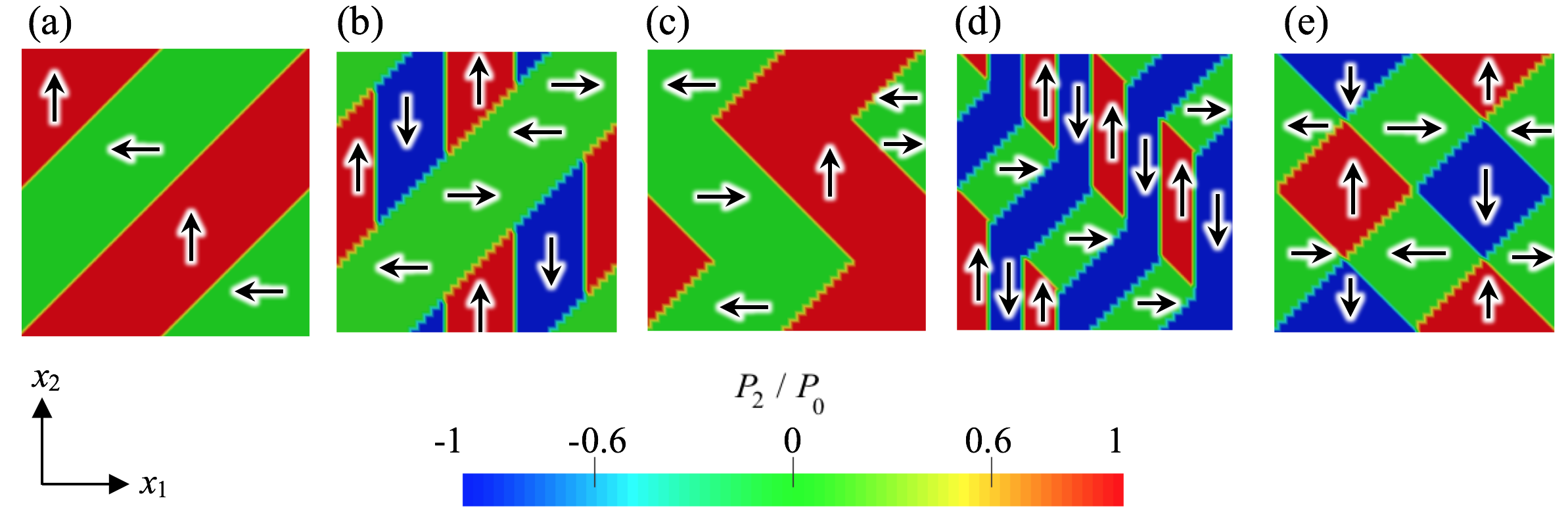}
\par\end{centering}
{\footnotesize{}\caption{{\small{}\label{fig:2}Establishing the stability of periodic domain
patterns (a) stripe (b) herringbone (c) herringbone-monodomain (d)
herringbone-stripes and (e) checkerboard, under stress-free and electric
field-free boundary conditions. $P_{0}=0.26\mathrm{C/m^{2}}$ is the
spontaneous polarization of BaTiO$_{3}$. }}
}{\footnotesize \par}
\end{figure}
\par\end{center}{\footnotesize \par}

\textcolor{black}{\small{}Next, the stability of the two candidate
patterns, herringbone and stripe domains, is explored under external
mechanical loading in the form of compressive strain, $\epsilon_{11}$
imposed by a substrate. The stability of the domain pattern under
an external strain field is a useful property for the energy harvester
device, which generates electrical energy from mechanical loads. Here,
a uniform strain $\epsilon_{11}$ is imposed by displacing the nodes
along $x_{2}=0$ of the periodic cell, while the boundary conditions
for nodes along $x_{1}=0$ and $x_{1}=L$ remain periodic, as given
by equation }{\small{}\ref{eq:3}}\textcolor{black}{\small{}. }{\small \par}
\begin{center}
{\footnotesize{}}
\begin{figure}
\begin{centering}
\includegraphics[width=0.6\columnwidth]{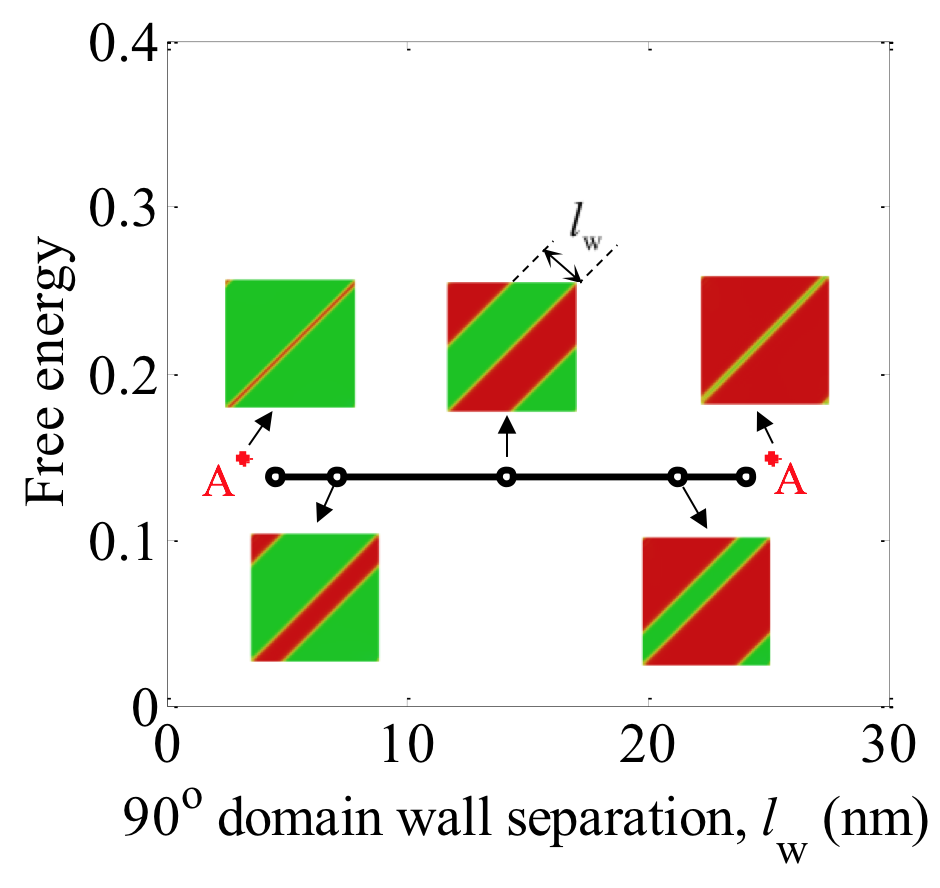}
\par\end{centering}
{\footnotesize{}\caption{{\small{}\label{fig:3}The volume average free energy (normalized
\cite{key-35}) of the stripe domain pattern as a function of domain
wall separation, $l_{w}$. Inset schematics denote polarization patterns.
The solid line connects the free energy of domain patterns at equilibrium,
while the markers \textquoteleft A\textquoteright{} at the far end
of the plots indicate the free energy of unstable domain patterns
before collapse into a monodomain at equilibrium. }}
}{\footnotesize \par}
\end{figure}
\par\end{center}{\footnotesize \par}

\textcolor{black}{\small{}A small axial strain $\epsilon_{11}=0.1\epsilon_{0}$,
where $\epsilon_{0}=0.0082$ is the spontaneous strain of BaTiO$_{3}$,
was found to destabilize the herringbone pattern causing the domains
to merge together at equilibrium. This pattern did not undergo cyclic
polarization switching and so appeared unsuitable for energy harvesting.
By contrast, the stripe domain pattern was stable over a wide range
of imposed strain, $0\leq\epsilon_{11}\leq0.65\epsilon_{0}$. The
width of the stripes in this domain pattern changes in proportion
to the applied strain. However the overall domain configuration remains
the same; this aspect is discussed in detail in section 4. This feature
of cyclic switching with domain walls sweeping back and forth but
remaining in the same pattern appears promising for energy harvesting.
Previous works indicate that applied stresses \cite{key-36new} and
imposed substrate strains \cite{key-37new} initiate the $90^{\circ}$
ferroelactic domain wall movement in polarization patterns. We next
explore the effect of domain wall position within the stripe domain
pattern by changing the separation distance $l_{w}$ of adjacent domain
walls, while keeping the size of the periodic cell constant. Here,
we initialize the stripe domain pattern with varying separation distance
$l_{w}$, under zero external load conditions, equation }{\small{}\ref{eq:3}}\textcolor{black}{\small{},
and relax towards equilibrium. The volume average free energy is plotted
as a function of separation distance in figure \ref{fig:3}. The relaxation
process produces no significant change in the domain pattern, apart
from smoothing the domain walls. The free energy is almost independent
of the domain wall separation for the stripe pattern, indicating that
the domain walls are in a neutral equilibrium. When $l_{w}\sim4\mathrm{nm}$,
the pattern is unstable and collapses to a monodomain. A symmetrically
opposite state has $l_{w}\sim25\mathrm{nm}$ and also collapses. In
these cases, the energy was calculated before collapse (not in equilibrium)
see markers \textquoteleft A\textquoteright{} at far ends in figure
\ref{fig:3}. The key result is that the stripe pattern is stable
over a wide range of separation of domain walls and also under external
mechanical loads, which motivates us to employ this pattern for designing
an energy harvester concept. }{\small \par}
\begin{singlespace}

\section*{{\small{}Energy harvester concept}}
\end{singlespace}

\noindent \textcolor{black}{\small{}The proposed energy harvester
concept is a device that cyclically converts input mechanical energy,
in the form of stresses/substrate strains, into electrical energy.
The concept is similar to existing designs of piezoelectric vibration
energy harvesters, except that the device undergoes ferroelectric/ferroelastic
switching during each cycle of operation. The harvester comprises
a barium titanate thin film on a conductive substrate with a top electrode,
see figure \ref{fig:4}(a-c). }{\small \par}
\begin{center}
{\footnotesize{}}
\begin{figure}
\begin{centering}
\includegraphics[width=1\columnwidth]{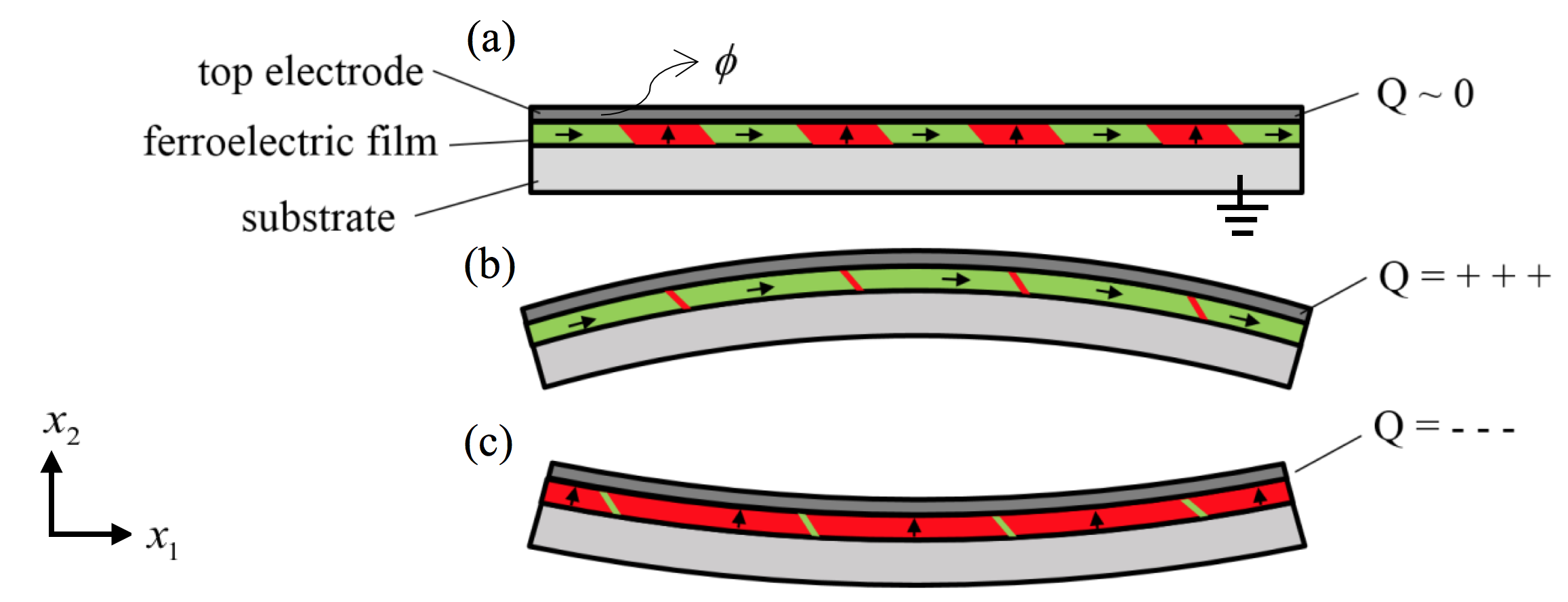}
\par\end{centering}
{\footnotesize{}\caption{{\small{}\label{fig:4}Schematic illustration of the working concept
of the energy harvester}}
}{\footnotesize \par}
\end{figure}
\par\end{center}{\footnotesize \par}

\noindent \textcolor{black}{\small{}The top electrode is connected
to an external circuit at a voltage $\phi$, while the substrate is
electrically grounded. The initial state is taken to be a stripe domain
pattern with zero net charge on the top electrode (this is an arbitrary
reference state for surface charge density). On bending the substrate,
a mechanical strain is induced in the ferroelectric film, causing
the polarized domains to switch. For example, in figure \ref{fig:4}(b),
tensile strain along the length of the ferroelectric film is expected
to cause the domains with polarization aligned along the $x_{1}-$
axis to grow in size; conversely, compressing the film by reversing
the sense of bending, see figure \ref{fig:4}(c), causes the domains
polarized along the $x_{2}-$ axis to grow in size. This change in
domain sizes drives a net electric charge onto or off the top electrode,
enabling electrical work to be done in an external circuit. Now, if
voltage $\phi$ is maintained constant throughout the cycle, then
charge flows onto and off the top electrode at the same voltage and
no net electrical work is done. Instead, suppose that during the compressive
part of the cycle, voltage $\phi=0$ so that charge flows off the
electrode without electrical work, and then $\phi$ is switched to
a negative value for the tensile part of the cycle. In this case the
tensile part of the cycle drives charge against a negative voltage
and electrical work can be extracted. Such an operation requires external
energy harvesting circuitry that will switch the electrode voltage
in phase with the mechanical vibration cycle. We return to this aspect
later in this section. }{\small \par}

\textcolor{black}{\small{}To model the energy harvester concept, a
square periodic cell of BaTiO$_{3}$ is simulated with plane strain
and electric field conditions, see figure \ref{fig:5}. A cell with
side $L=40$nm is chosen, noting that at this size the domain spacing
in the stripe domain pattern is about $l_{w}=14\mathrm{nm}$ and the
domain walls have freedom to move without the pattern collapsing into
a monodomain. An axial in-plane strain, $\epsilon_{11}$ is induced
by the conductive substrate and modelled by imposing nodal displacements
along the edge AB at $x_{2}=0$. At the top surface, $x_{2}=L$ an
electrode of area $A$ is modelled by imposing voltage $\phi$; for
simplicity it is assumed that the electrode does not provide significant
mechanical constraint. Nodes along $x_{1}=0$ and $x_{1}=L$ are constrained
with the periodic boundary conditions given by equation }{\small{}\ref{eq:3}}\textcolor{black}{\small{}.
The simulation is initialized by imposing a stripe pattern on the
periodic cell with sharp interfaces. The voltage on the top electrode
is set to $\phi=0$ and the initially imposed substrate strain $\epsilon_{11}=0.3\epsilon_{0}$
is applied at $x_{2}=0$. This choice of substrate strain matches
the average strain of the unstressed stripe pattern with equal volume
fractions of the two domain types, found in prior calculations. The
periodic cell of the energy harvester is then solved using the phase-field
model to find an equilibrium stripe-like pattern, see figure \ref{fig:6}(a).
At $\epsilon_{11}=0.3\epsilon_{0}$ the domains in the periodic cell,
that are polarized along the $x_{1}-$ axis ($P_{1}$ domains) and
the $x_{2}-$ axis ($P_{2}$ domains) are of equal sizes and a net
charge of $-0.13\mathrm{C/m^{2}}$ is attracted onto the top electrode. }{\small \par}
\begin{center}
{\footnotesize{}}
\begin{figure}
\begin{centering}
\includegraphics[width=0.9\columnwidth]{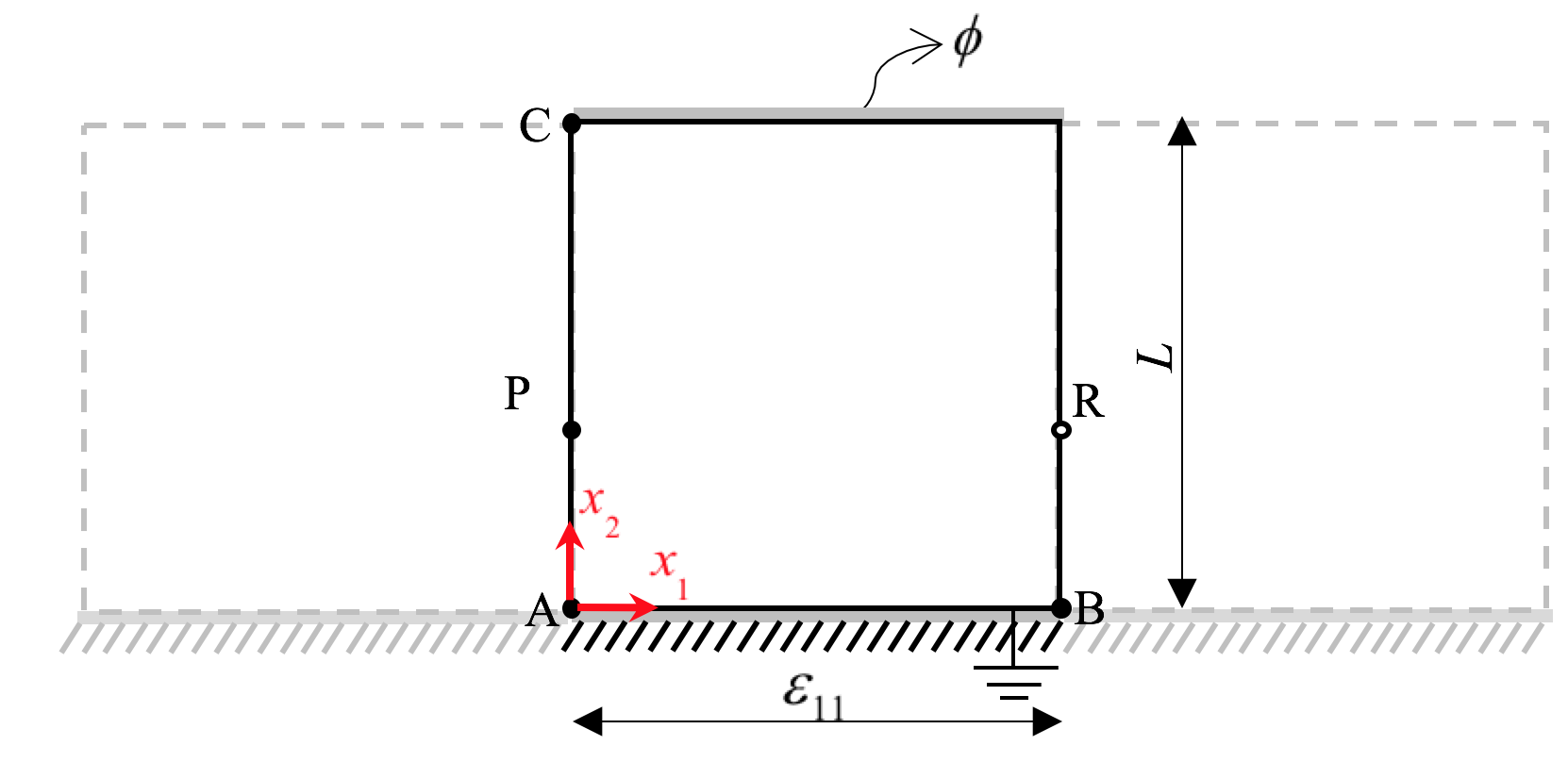}
\par\end{centering}
{\footnotesize{}\caption{{\small{}\label{fig:5}Schematic illustration of the boundary conditions
on a periodic cell of the energy harvester.}}
}{\footnotesize \par}
\end{figure}
\par\end{center}{\footnotesize \par}

\textcolor{black}{\small{}In order to achieve cyclic operation, it
is necessary to limit the strain imposed by the substrate such that
the film remains configured as a stripe pattern. If the substrate
is strained too far, either in tension or in compression, a single
domain state will be produced and the imposed periodicity of the pattern
would be lost. To explore the tolerable limits of straining, a range
of substrate strains was imposed while maintaining $\phi=0$. Starting
from the state with equal domain fractions, mechanical loading was
applied to the energy harvester by incrementing the substrate strain
in steps of $0.05\epsilon_{0}$. In the first step the equilibrium
domain pattern formed at $\epsilon_{11}=0.3\epsilon_{0}$ is taken
as the initial state and the substrate strain is increased to $0.35\epsilon_{0}$.
The system is then allowed to attain equilibrium. Subsequent steps
increment or decrement the substrate strain in the range $-0.1\epsilon_{0}\leq\epsilon_{11}\leq0.7\epsilon_{0}$,
finding an equilibrium state at each step. The net charge per unit
area collected on the top electrode is computed at each value of $\epsilon_{11}$
(with $\phi=0$) and the results are shown in figure \ref{fig:6}(b). }{\small \par}
\begin{center}
{\footnotesize{}}
\begin{figure}
\begin{centering}
\includegraphics[width=1\columnwidth]{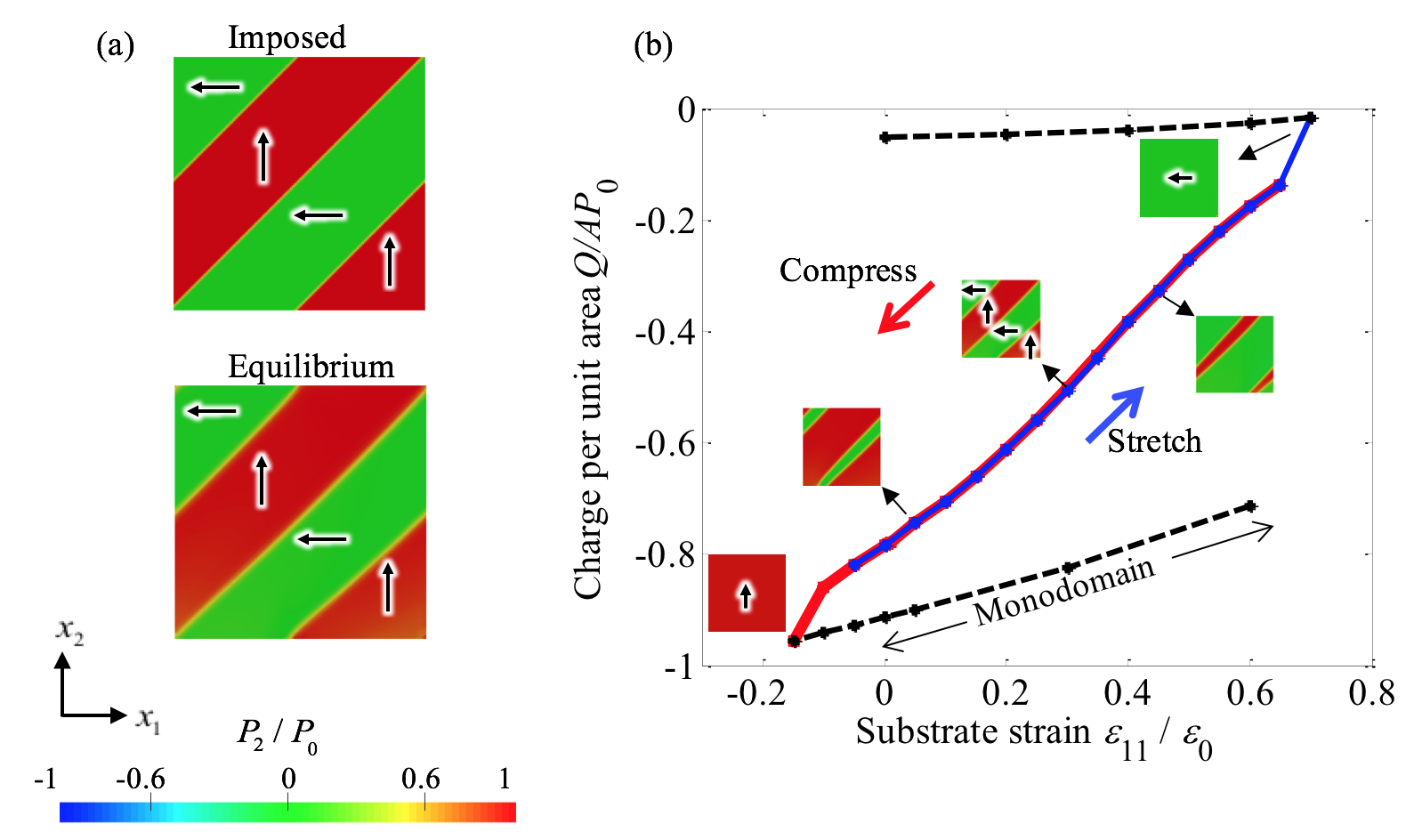}
\par\end{centering}
{\footnotesize{}\caption{{\small{}\label{fig:6}(a) Polarization pattern in the initial state
of the energy harvester at $\epsilon_{11}=0.3\epsilon_{0}$. (b) Charge
per unit area collected on the top electrode as a function of the
substrate strain in the energy harvester. $A$ is the area of the
top electrode and $P_{0}=0.26\mathrm{C/m^{2}}$. The cycle is performed
at $\phi=0$ and inset schematics illustrate the domain patterns polarized
along the  $x_{2}-$ axis. }}
}{\footnotesize \par}
\end{figure}
\par\end{center}{\footnotesize \par}

\textcolor{black}{\small{}It can be seen from figure \ref{fig:6}(b)
that increasing the strain makes the net charge on the top electrode
less negative. This is because the sizes of $P_{2}$ domains decrease
with increasing $\epsilon_{11}$, see inset schematics in figure \ref{fig:6}(b).
On compressing the substrate, the net charge on the top electrode
becomes more negative. Significantly, for tensile substrate strains,
$\epsilon_{11}>0.65\epsilon_{0}$, or for compressive strains $\epsilon_{11}<-0.1\epsilon_{0}$,
the stripe-like features in the periodic cell disappear leading to
the formation of a uniformly polarized monodomain. Upon stretching/compressing
the monodomain, the stripe-like features are not recovered, and the
cycle is not repeatable. However, provided the substrate strain is
limited to the range $0\leq\epsilon_{11}\leq0.65\epsilon_{0}$ a stable
cycle occurs with domain walls sweeping back and forth through the
film. }{\small \par}

\textcolor{black}{\small{}Next, the challenge is to extract electrical
energy from a cycle of mechanical straining of the energy harvester.
In order to achieve this, the voltage on the top electrode is varied
during the energy harvester cycle, allowing electrical work to be
extracted. An example of the energy harvester cycle is given in four
steps, see table 1. The corresponding phase-field simulations are
illustrated in figure \ref{fig:7}. }{\small \par}
\begin{center}
\begin{table}
\noindent %
\begin{tabular*}{1\columnwidth}{@{\extracolsep{\fill}}>{\raggedright}p{0.04\columnwidth}>{\raggedright}p{0.2\columnwidth}>{\raggedright}p{0.25\columnwidth}>{\raggedright}p{0.4\columnwidth}}
\toprule 
{\scriptsize{}Step} &
{\scriptsize{}Voltage } &
{\scriptsize{}Substrate strain } &
{\scriptsize{}Charge per unit area }\tabularnewline
\midrule
{\scriptsize{}AB} &
{\scriptsize{}Constant at $\phi=0$} &
{\scriptsize{}Decreased $\epsilon_{11}=0.65\epsilon_{0}\rightarrow0$} &
{\scriptsize{}$Q/A=-0.2P_{0}\rightarrow-0.8P_{0}$}\tabularnewline
{\scriptsize{}BC} &
{\scriptsize{}Decreased to $\phi=-0.25\mathrm{V}$} &
{\scriptsize{}Constant at $\epsilon_{11}=0$} &
{\scriptsize{}$Q/A=-0.8P_{0}\rightarrow-0.9P_{0}$}\tabularnewline
{\scriptsize{}CD} &
{\scriptsize{}Constant at $\phi=-0.25\mathrm{V}$} &
{\scriptsize{}Increased $\epsilon=0\rightarrow0.65\epsilon_{0}$} &
{\scriptsize{}$Q/A=-0.9P_{0}\rightarrow-0.35P_{0}$}\tabularnewline
{\scriptsize{}DA} &
{\scriptsize{}Increased to $\phi=0$} &
{\scriptsize{}Constant at $\epsilon_{11}=0.65\epsilon_{0}$} &
{\scriptsize{}$Q/A=-0.35P_{0}\rightarrow-0.2P_{0}$}\tabularnewline
\bottomrule
\end{tabular*}

\caption{{\small{}\label{Tab:1}}\textcolor{black}{\small{}Sequential steps
in the energy harvester cycle. The steps A-B-C-D correspond to the
plot in figure \ref{fig:8}(a-b). $Q/A$ is the net charge per unit
area on the top electrode. }}
\end{table}
\par\end{center}

\textcolor{black}{\small{}The domain evolution through step AB, is
illustrated by figure \ref{fig:7}(a-d). Here, the voltage on the
top electrode is held constant at $\phi=0$, while the substrate strain
is decreased in steps of $0.05\epsilon_{0}$. The $P_{2}$ domains
grow and the charge on the top electrode changes as shown in figure \ref{fig:8}(a).
The portions of the domain walls in close proximity to the substrate
curve due to the imposed strain, see figure \ref{fig:7}(b). At point
B, $\phi=0$ and $\epsilon_{11}=0$, while the net charge per unit
area on the top electrode is $\sim-0.8P_{0}$. Now, the substrate
strain is briefly held constant at zero strain while the voltage on
the top electrode is switched to $-0.25\mathrm{V}$, step BC. During
this step, further net charge per unit area of $\sim-0.1P_{0}$ flows
on to the top electrode, and the domains polarized along  $x_{2}-$
axis grow slightly, see figure \ref{fig:7}(d-e). In the subsequent
step CD, the voltage on the top electrode is held constant at $-0.25\mathrm{V}$
and the substrate is stretched. Now the domains polarized along the
direction of applied axial strain increase in size, see figure \ref{fig:7}(f-i)
changing the electrode charge to $\sim-0.35P_{0}$. At point D, with
$\epsilon_{11}=0.65\epsilon_{0}$, the substrate strain is again briefly
held constant while the voltage switches back to 0V. This causes the
charge per unit area to return to its initial state $\sim-0.2P_{0}$,
see also figure \ref{fig:7}(i-j). The energy harvester has now returned
to its initial state, and the cycle can be repeated. The electrical
work output per unit area in this cycle is $\sim0.04\mathrm{J/m^{2}}$. }{\small \par}
\begin{center}
{\footnotesize{}}
\begin{figure}
\begin{centering}
\includegraphics[width=1\columnwidth]{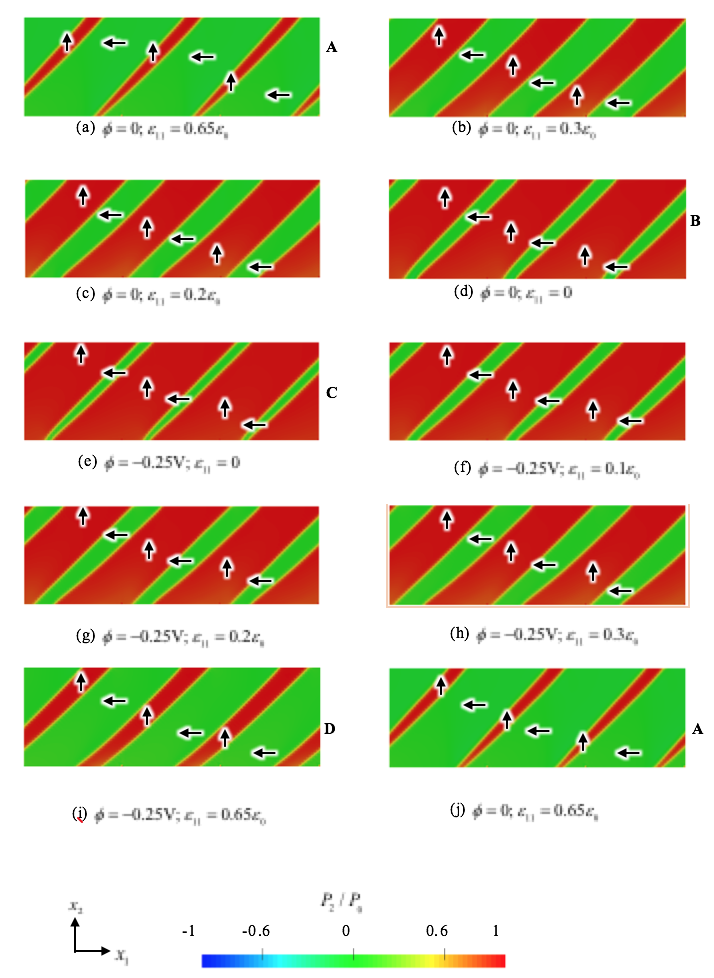}
\par\end{centering}
{\footnotesize{}\caption{{\small{}\label{fig:7}Domain patterns illustrating the working cycle
of a conceptual energy harvester. }}
}{\footnotesize \par}
\end{figure}
\par\end{center}{\footnotesize \par}

\begin{center}
{\footnotesize{}}
\begin{figure}
\begin{centering}
\includegraphics[width=1\columnwidth]{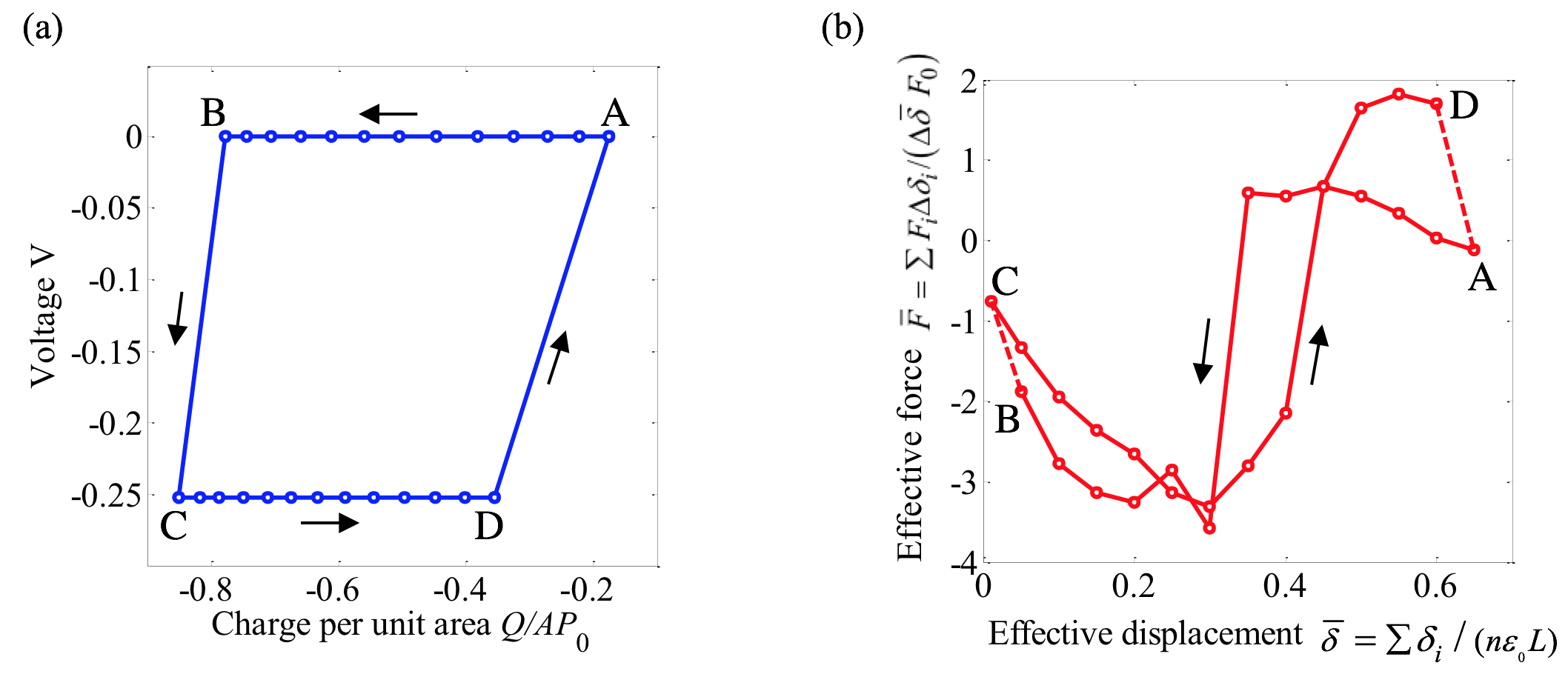}
\par\end{centering}
{\footnotesize{}\caption{{\small{}\label{fig:8}Plot of (a) voltage-charge (b) effective force-displacement
in the energy harvester cycle shown in figure }\textcolor{black}{\small{}\ref{fig:7}}{\small{}.
$F_{0}=0.692\mathrm{N/m}$.}}
}{\footnotesize \par}
\end{figure}
\par\end{center}{\footnotesize \par}

\textcolor{black}{\small{}An effective force-displacement curve for
the energy harvester is plotted in figure \ref{fig:8}(b). Here the
effective displacement $\bar{\delta}$ is defined as a normalized
average over the nodes on the substrate, $\bar{\delta}=\Sigma\delta_{i}/(n\epsilon_{0}L)$,
where $\delta_{i}$ is the displacement of the $i$th node and $n$
is the total number of substrate nodes. An effective force at the
end of each step in the phase-field simulation is calculated by $\bar{F}=\Sigma F_{i}\Delta\delta_{i}/\Delta\bar{\delta}F_{0}$,
where $F_{i}$ is the nodal force value and the normalization for
force per unit length is $F_{0}=0.692\mathrm{N/m}$. This definition
gives a work equivalent force such that the mechanical work increment
in each step is $\bar{F}\Delta\bar{\delta}$. In figure \ref{fig:8}(b),
sharp transitions in effective force are observed on the curve segment
AB upon decreasing $\bar{\delta}=0.35\rightarrow0.30$, and likewise
on the segment CD curve on increasing $\bar{\delta}=0.40\rightarrow0.45$.
These sharp transitions are caused by domain wall movements or polarization
switching, $-P_{1}\rightarrow P_{2}$ or $P_{2}\rightarrow-P_{1}$,
during the energy harvester cycle. Polarization switching is accompanied
by a change in spontaneous strain values that affects the nodal forces.
This abrupt change in nodal force is observed on each substrate node
when a domain wall sweeps across it. The shape of the effective force-displacement
curve is complicated. In figure \ref{fig:8}(b), the primary contribution
to the effective force, $\bar{F}$ arises from nodes closer to the
edge of the periodic cell as $x_{1}\rightarrow L$. This is because,
for a given substrate strain, $\epsilon_{11}$ the the $i$th node
is displaced as $\Delta\delta_{i}=\epsilon_{11}x_{i}$. }{\small \par}

\textcolor{black}{\small{}$\Delta\delta_{i}$ is thus greatest for
nodes near $x_{1}\rightarrow L$, and least for nodes at $x_{1}\rightarrow0$.
Figure \ref{fig:9} shows the nodal force and displacement values
at the node $(x_{1},x_{2})=(40\mathrm{nm,}0)$. Sharp transitions
observed in figure \ref{fig:8}(b) correspond with the changes in
figure \ref{fig:9}, which is caused by a domain wall sweeping across
the corner of the periodic cell at $(x_{1},x_{2})=(40\mathrm{nm,}0)$. }{\small \par}
\begin{center}
{\footnotesize{}}
\begin{figure}
\begin{centering}
\includegraphics[width=0.7\columnwidth]{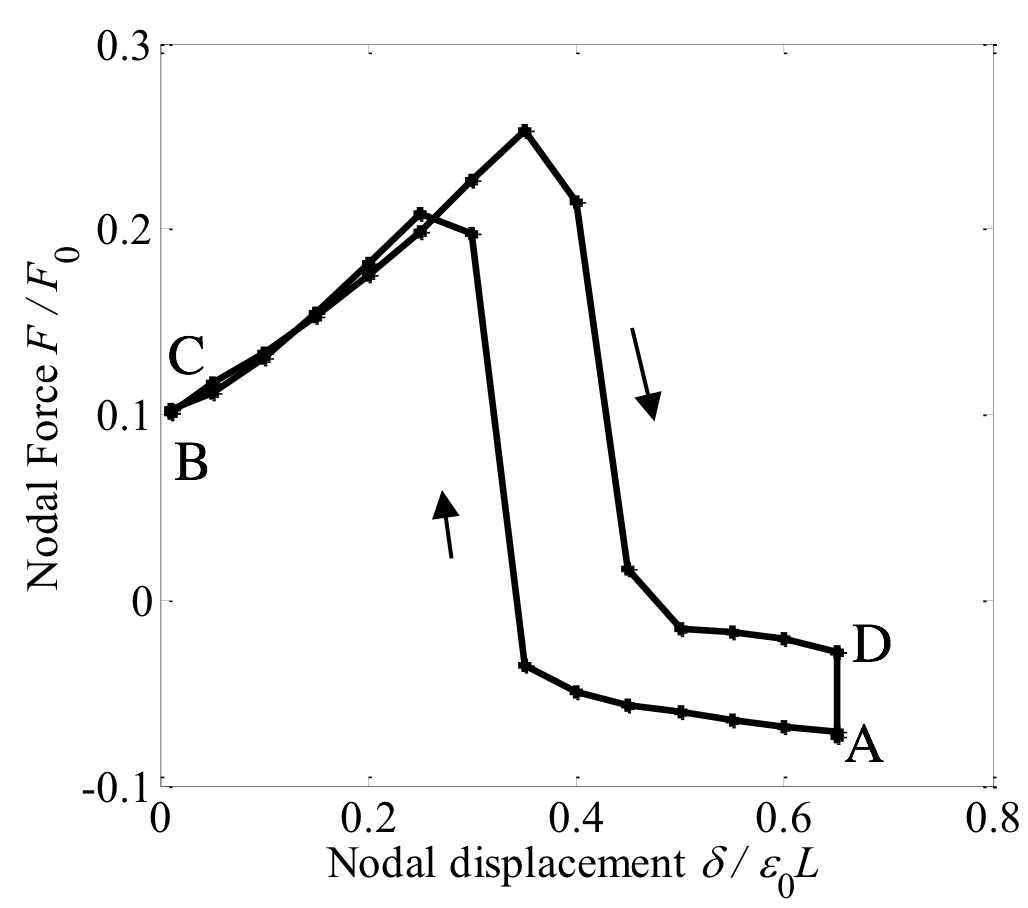}
\par\end{centering}
{\footnotesize{}\caption{{\small{}\label{fig:9}Plot of nodal forces and displacements on \textquoteleft S\textquoteright{}
at }\textcolor{black}{\small{}$(x_{1},x_{2})=(40\mathrm{nm,}0)$}{\small{}
of the periodic cell, during the energy harvester cycle. Labels A-D
corresponds to the harvester states in figure \ref{fig:7}. }}
}{\footnotesize \par}
\end{figure}
\par\end{center}{\footnotesize \par}

\textcolor{black}{\small{}The net electrical work output, from the
voltage-charge curve, is $\sim0.04\mathrm{J/m^{2}}$ in each cycle.
If operated at frequency 1kHz, a power density of $~40\mathrm{W/m^{2}}$
could be potentially generated, assuming ideal conditions. Since the
external harvesting circuit in this design is required to carry out
voltage switching at a frequency matching the driving oscillations,
it is anticipated that the device would be designed with a macroscopic
resonance frequency matching the available mechanical energy source.
Suitable harvesting circuits have been designed for similar applications
\cite{key-36,key-37}. Limitation of the substrate strain to safe
operating levels could be practically achieved by blocking macroscopic
oscillations greater than the limiting amplitude. }{\small \par}

\textcolor{black}{\small{}It is of interest to compare the power density
achieved by the proposed device with that of other energy harvesting
technologies. A wide range of mechanical energy harvesters has been
developed in recent years, including piezoelectrics \cite{key-18,key-38,key-39,key-40},
nanowire arrays \cite{key-41,key-42,key-43,key-44,key-45,key-46,key-47},
triboelectric devices \cite{key-48,key-49,key-50} and magnetostrictives
\cite{key-51,key-52,key-53}. Power densities are commonly expressed
in terms of power per unit volume of working material at an operating
frequency. Alternatively, for comparison with area based technologies
such as photovoltaics, an area power density can be derived by factoring
in the thickness of working material. Considering the well-established
piezoelectric energy harvesters, cantilever devices such as that of
Shen }\textit{\textcolor{black}{\small{}et al.}}\textcolor{black}{\small{}
\cite{key-40} typically offer area power density of $0.2-0.3\mathrm{W/m^{2}}$
at around 200Hz. At this frequency, the proposed device could generate
around $7\mathrm{W/m^{2}}$. The enhancement of more than an order
of magnitude is due to the use of ferroelectric switching in place
of piezoelectric response. Devices based on the bending of a nanowire
array \cite{key-41,key-42,key-43,key-44,key-45,key-46,key-47} appear
to result in a wide range of power densities depending on the choice
of material and operating regime. At the upper end of performance
\cite{key-43,key-44,key-46} the power density is comparable to that
of the proposed ferroelectric harvester. The power generated by energy
harvesters based on the concept of triboelectricity that uses contact
electrification between the thin films of a polymer and a metal foil
\cite{key-48} can reaches up to $30-40\mathrm{W/m^{2}}$ at an operating
frequency of 10Hz. This power density is about 100 times greater than
the thin film ferroelectric energy harvester concept operating at
the same frequency. By contrast, magnetostrictive energy harvesters
such as that of Wang and Yuan \cite{key-51} can achieve power densities
around tens of mW/m$^{2}$, and so are better suited to applications
requiring low power density. Finally it is worth noting that ground
based solar photovoltaics which offer $\sim10\%$ energy conversion
efficiency \cite{key-54,key-55,key-56}, commonly achieve power densities
of around $30-50\mathrm{W/m^{2}}$, which is similar to the proposed
ferroelectric device operating at 1kHz. }{\small \par}
\begin{singlespace}

\section*{{\small{}Conclusion}}
\end{singlespace}

\noindent \textcolor{black}{\small{}A phase-field model has been employed
to identify stable domain patterns and to explore the concept of an
energy harvester device. A stripe domain pattern with in-plane polarizations
was found to be stable over a wide range of externally imposed strains
and its potential application in a conceptual thin film energy harvester
device was demonstrated. The results indicate the energy harvester
generates about 0.04J/m$^{2}$ on each cycle. The device thus has
power generation capability that compares favourably with a range
of energy harvesting technologies, provided sufficient vibration energy
is available at its operating frequency. The work also illustrates
a potential use of the phase-field model as a design tool in domain
engineering of nanoscale devices.}{\small \par}
\begin{singlespace}

\section*{{\small{}Acknowledgements}}
\end{singlespace}

\noindent \textcolor{black}{\small{}The authors wish to thank Professor
C. M. Landis for help in providing program codes and advice. A. Renuka
Balakrishna gratefully acknowledges support of the Felix scholarship
trust and British Federation of Women Graduates. }{\small \par}

\noindent %

\end{document}